\documentclass[12pt]{iopart}
\usepackage{graphicx}
\usepackage{epstopdf}

\begin{document}

\title{Magnetic and electronic properties of La$_3M$O$_7$ and possible polaron formation in hole-doped La$_3M$O$_7$ ($M$=Ru and Os)}

\author{Bin Gao, Yakui Weng, Jun-Jie Zhang, Huimin Zhang, Yang Zhang and Shuai Dong}
\ead{sdong@seu.edu.cn}
\address{Department of Physics, Southeast University, Nanjing 211189, China}
\vspace{2pc}

\begin{abstract}
Oxides with $4d$/$5d$ transition metal ions are physically interesting for their particular crystalline structures as well as the spin-orbit coupled electronic structures. Recent experiments revealed a series of $4d$/$5d$ transition metal oxides $R_3M$O$_7$ ($R$: rare earth; $M$: $4d$/$5d$ transition metal) with unique quasi-one-dimensional $M$ chains. Here first-principles calculations have been performed to study the electronic structures of La$_3$OsO$_7$ and La$_3$RuO$_7$. Our study confirm both of them to be Mott insulating antiferromagnets with identical magnetic order. The reduced magnetic moments, which are much smaller than the expected value for ideal high-spin state ($3$ $t_{2g}$ orbitals occupied), are attributed to the strong $p-d$ hybridization with oxygen ions, instead of the spin-orbit coupling. The Ca-doping to La$_3$OsO$_7$ and La$_3$RuO$_7$ can not only modulate the nominal carrier density but also affect the orbital order as well as the local distortions. The Coulombic attraction and particular orbital order would prefer to form polarons, which might explain the puzzling insulating behavior of doped $5d$ transition metal oxides. In addition, our calculation predict that the Ca-doping can trigger ferromagnetism in La$_3$RuO$_7$ but not in La$_3$OsO$_7$.
\end{abstract}


\section{Introduction}
Transition metal oxides have attracted enormous attentions for its plethoric members, divergent properties, novel physics, and great impacts on potential applications based on correlated electrons. In past decades, the overwhelming balance of interests were devoted to those compounds with $3d$ elements, which showed high-$T_{\rm C}$ superconductivity, colossal magnetoresistivity, multiferroicity, and so on~\cite{Dagotto:Sci,Dong:Ap}. However, the $4d$ and $5d$ counterparts were much less concerned and only in very recent years a few of them, e.g. Sr$_2$IrO$_4$, have been focused on~\cite{Kim:Sci,Meng:prl,Kim:Prl5}. In principle, for $4d$/$5d$ electrons, the electron-electron repulsion, e.g. Hubbard $U$, is much weaker due to more extended wave functions, while the spin-orbit coupling (SOC) is much stronger due to the large atom number, comparing with the $3d$ electrons~\cite{Cao:Bok}. These characters may lead to non-conventional physics in $4d$/$5d$ metal oxides, e.g. $p$-wave superconductors, spin-orbit Mott insulator, Kitaev magnets, topological materials, and possible high-$T_{\rm C}$ superconductors~\cite{Kim:Prl5,Nelson:Sci,Kitaev:Anp,Witczak-Krempa:Arcmp,Wang:Prl}.

\begin{figure}
\centering
\includegraphics[width=0.68\textwidth]{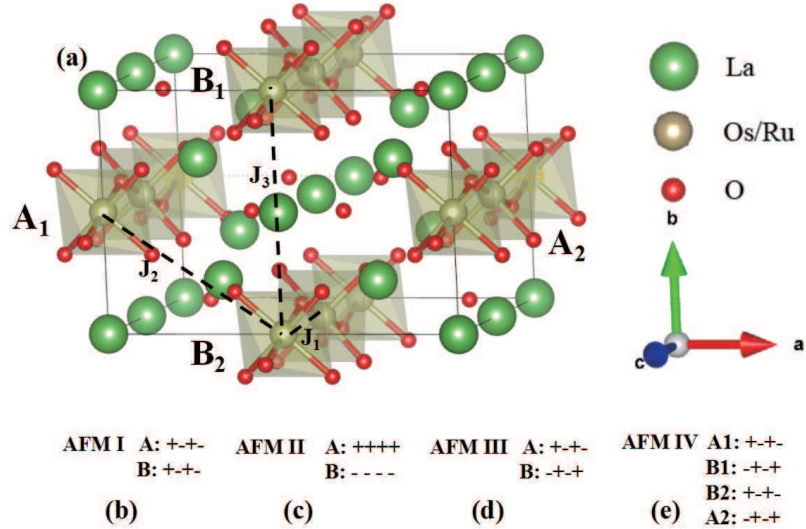}
\caption{Structure and possible magnetism of La$_3M$O$_7$ ($M$=Os or Ru). (a) Schematic of the weberite $R_3M$O$_7$ structure. (b)-(e) The possible AFM orders considered in our calculations. A and B are the chain indexes of $M$ chains. $+$ and $-$ denote the spin directions of $M$.}
\label{Fig1}
\end{figure}

Till now, the most studied $4d$/$5d$ metal oxides owns quasi-two-dimensional layer structures (e.g. Sr$_2$IrO$_4$ and Na$_2$IrO$_3$) or three-dimensional structures (e.g. SrIrO$_3$ and SrRuO$_3$). Recently, those $4d$/$5d$ metal oxides with quasi-one-dimensional chains have also been synthesized, which may lead to unique low-dimensional physics, e.g. charge density waves, spin-Peierls transitions, and novel magnetic excitations~\cite{Hase:Prl,Monceau:Prl,Tennant:Prb}. For example, recent experiments reported the basic physical properties of $R_3M$O$_7$, which owns the weberite structure, as shown in Fig.~\ref{Fig1}(a)~\cite{Allpress:Jssc}. Since here the $4d$/$5d$ electrons are mostly confined in one-dimensional chains instead of two-dimensional plane or three-dimensional framework, their electronic and magnetic structures, may be markedly different from the higher-dimensional structural $4d$/$5d$ counterparts. Given the decreased electron correlations and increased SOC of the $4d$/$5d$ electrons, the physical behavior of these compounds may also show differences comparing with quasi-one-dimensional $3d$ metal oxides~\cite{Comini:Pms}. In fact, there is rare $3d$ metal oxide forming the weberite $R_3M$O$_7$ structure. It is therefore physically interest to study these new systems.

Taking La$_3$OsO$_7$ for example, recent experimental studies reported its structural, transport, and magnetic properties, characterized by magnetic susceptibility, x-ray diffraction, as well as neutron diffraction~\cite{Morrow:Prb}. The corner-shared OsO$_6$ octahedra form chains along the [001] direction of the orthorhombic framework. The nearest-neighbor distance of Os-Os is $3.81$ {\AA} within a chain, but $6.75$ {\AA} between chains. The Os-O-Os bond angle within a chain is about $153^\circ$, implying strong octahedra tilting, which is also widely observed in other oxides. Its ground state is an antiferromagnetic (AFM) insulator. The Ca-doped La$_3$OsO$_7$ was also studied. Despite the change of nominal carrier density, surprisingly, this hope-doped system remain an insulator (or a semiconductor), violating the rigid band scenario~\cite{Morrow:Prb}. Similar robust insulating behavior was also found in some doped iridates~\cite{Wang:Prb6,Lu:Prb6}, which was expected to show superconductivity after doping~\cite{Watanabe:Prl,Wang:Prl}.

In this work, we have performed systematic first-principles calculations to understand the electronic and magnetic properties of La$_3$OsO$_7$ as well as the isostructural La$_3$RuO$_7$. The doping effect has also been studied, which may provide a reasonable explanation to the insulating behavior, based on the polaron forming. To our best knowledge, there were very few theoretical studies on these two materials before. Only Khalifah {\it et al.} calculated several magnetic states of La$_3$RuO$_7$~\cite{Khalifah:Prb1}. Even though, their predicted ground state (see Fig.~\ref{Fig1}(b)) seems to be inaccurate, according to our results.

\section{Model \& methods}
All following calculations were performed using the Vienna {\it ab initio} Simulation Package (VASP) based on the generalized gradient approximation (GGA)~\cite{Kresse:Prb,Kresse:Prb96}. The new-developed PBEsol function is adopted~\cite{Perdew:Prl3}, which can improve the accurate description of crystal structure comparing with the old-fashion PBE one. The plane-wave cutoff is $550$ eV and Monkhorst-Pack $k$-points mesh centered at $\Gamma$ points is adopted.

Starting from the low-temperature experimental orthorhombic (No. $63$ $Cmcm$) structures~\cite{Morrow:Prb,Lam:Jssc}, the lattice constants and inner atomic positions are fully optimized till the Hellman-Feynman forces are all less than $0.01$ eV/{\AA}. The Hubbard repulsion $U_{\rm eff}$ ($=U-J$) is imposed on Ru's $4d$ orbitals and Os's $5d$ orbitals~\cite{Dudarev:Prb}. Various values of $U_{\rm eff}$ have been tested from $0$ eV to $4$ eV. It is found that $U_{\rm eff}$(Ru)=$1$ eV is the best choice to reproduce the experimental structure of La$_3$RuO$_7$, while for La$_3$OsO$_7$ the bare GGA without $U_{\rm eff}$ is the best choice. Comparing with experimental values~\cite{Morrow:Prb,Khalifah:Prb1}, the deviation of calculated lattice constants are only $<0.8\%$ for La$_3$OsO$_7$, and $0.5\%$ for La$_3$RuO$_7$, providing a good start point to study other physical properties. These choices of $U_{\rm eff}$ are quite reasonable considering the gradually deceasing Hubbard repulsion from $3d$ to $5d$.

Considering the fact of heavy atoms, the relativistic SOC is also taken into consideration, comparing with those calculations without SOC.

\section{Results \& discussion}
\subsection{Undoped La$_3M$O$_7$: magnetic orders and reduced moments}
First, the magnetic ground state is checked by comparing several possible magnetic orders, including ferromagnetic (FM) state, and various AFM ones (AFM I-IV as shown in Fig.~\ref{Fig1}(b-e)). For AFM I, III, and IV states, the -up-down-up-down- ordering is adopted within each chain, but with different coupling between chains. Taking the FM state as the energy reference, the energies of all candidates are summarized in Table 1, which suggest the AFM IV to be the possible ground state for both $M$=Os and Ru.

\begin{table}
\centering
\caption{The energy difference (in unite of meV) for a minimal unit cell (four f.u.'s), local magnetic moment per $M$ within the default Wigner-Seitz sphere (in unit of $\mu_{\rm B}$), and band gap (in unit of eV) of La$_3M$O$_7$.}
\begin{tabular*}{0.68\textwidth}{@{\extracolsep{\fill}}llrcc}
\hline
\hline
 $~$ & Magnetism & Energy & Moment & Gap\\
\hline
La$_3$OsO$_7$ &      FM &     0 &  0.851,0.846 & 0.19 \\
              &   AFM I &  -689 & 1.670,-1.670 &  0.42 \\
              &  AFM II &  -228 & 1.435,-1.435 &     0 \\
              & AFM III &  -709 & 1.652,-1.652 &  0.54 \\
              &  AFM IV &  -714 & 1.651,-1.651 &  0.53 \\
\hline
La$_3$RuO$_7$ &      FM &     0 &  1.958,1.959 &  0.53 \\
              &   AFM I &   -23 & 1.902,-1.902 &  0.60 \\
              &  AFM II &   -22 & 1.948,-1.948 &  0.78 \\
              & AFM III &   -38 & 1.902,-1.902 &  0.73 \\
              &  AFM IV &   -42 & 1.900,-1.899 &  0.70 \\
\hline
\hline
\end{tabular*}
\label{Table 1}
\end{table}

By mapping the system to a classical spin model, the exchange coefficients between neighbor (within each chain and between chains, as indicated in Fig.~\ref{Fig1}(a)) spins (normalized to $|S|$=$1$) can be extracted as: $J_1$=$72.00$ meV, $J_2$=$8.34$ meV, and $J_3$=$2.96$ meV for $M$=Os, $J_1$=$3.45$ meV, $J_2$=$0.66$ meV, and $J_3$=$0.37$ meV for $M$=Ru. Obviously, the exchanges between Os chains are quite prominent even for the nearest neighbor chains (distance up to $6.75$ {\AA}), implying strongly coupled AFM chains, different from the one-dimensional intuition. And these exchanges are much stronger in La$_3$OsO$_7$ than the correspondences in La$_3$RuO$_7$. These characters of La$_3$OsO$_7$ are benefited from the more extended distribution of $5d$ orbitals.

Experimentally, the AFM transition temperatures of La$_3$OsO$_7$ are much higher than the corresponding ones of La$_3$RuO$_7$. For La$_3$OsO$_7$, the intrachain magnetic correlation emerges near $\sim100$ K (mainly due to $J_1$) and the fully three-dimensional AFM ordering occurs at $45$ K (also determined by $J_2$ and $J_3$~\cite{Khalifah:Prb1}. In contrast, the signal for magnetic ordering in La$_3$RuO$_7$ appears at $\sim17$ K with short-range characters~\cite{Khalifah:Prb1}. Note Ref.~\cite{Khalifah:Prb1} once predicted the ground state of La$_3$RuO$_7$ to be AFM I, which is ruled out according to our calculation. More neutron experiments are needed to refine the subtle magnetic order of La$_3$RuO$_7$.

\begin{figure}
\centering
\includegraphics[width=0.68\textwidth]{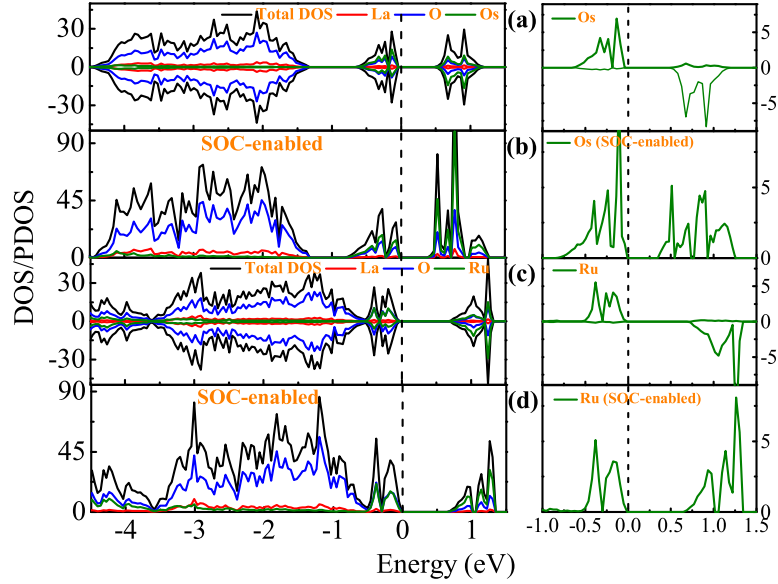}
\caption{Density of state (DOS) and projected density of state (PDOS) of La$_3M$O$_7$ ($M$=Os or Ru). (a-b) La$_3$OsO$_7$; (c-d) La$_3$RuO$_7$. (a) and (c) SOC-disabled; (b) and (d) SOC-enabled. Right panels: the corresponding near-Fermi-level PDOS of an individual $M$ ion.}
\label{Fig2}
\end{figure}

Second, the total density of states (DOS) and atomic-projected density of states (PDOS) of La$_3$OsO$_7$ are displayed in Fig.~\ref{Fig2}(a). Clearly, the system is insulating with a band gaps of $\sim0.53$ eV, even in the pure GGA calculation. Both the topmost valence band(s) and bottommost conducting band(s) of La$_3$OsO$_7$ are from Os, in particular the $t_{2g}$ orbitals. Since the $5d$ orbitals have a large SOC coefficient, we also calculate the DOS and PDOS with SOC enabled, which are presented in Fig.~\ref{Fig2}(b) for comparison. However, there is no qualitative difference between the SOC-enabled and SOC-disabled calculations. The quantitative differences include: 1) a shrunk band gap $\sim0.37$ eV (SOC-enabled); 2) a slightly reduced local magnetic moment from $1.661$ $\mu_{\rm B}$/Os (SOC-disabled) to $1.578$ $\mu_{\rm B}$/Os (SOC-enabled). In particular, the magnitude of orbital moment is only $\sim0.087$ $\mu_{\rm B}$. Noting this local moment is obtained by integrating the wave function within the Wigner-Seitz radius of Os ($0.58$ {\AA}) and thus not absolutely precise. Even though, the theoretical values are still quite close to the experimental one $\sim1.71$ $\mu_{\rm B}$/Os~\cite{Morrow:Prb}. Such a local moment is significantly reduced from the high-spin expectation ($3$ $\mu_{\rm B}$/Os) of three $t_{2g}$ electrons as in Os$^{5+}$ here, but agrees with recent neutron diffraction results of Os$^{5+}$ in several double perovskites~\cite{Kermarrec:Prb,Taylor:Prb,Kanungo:Prb}.

According to the PDOS (insert of Fig.~\ref{Fig2}(a-b)), every Os seems to be in the high-spin state, i.e. only spin-up electrons within the Wigner-Seitz radius. Then how to understand the reduced local moment? Above SOC-enabled calculation has ruled out SOC as the main contribution, which can only slightly affect the value of moment. Instead, the real mechanism is the covalency between Os and O. As revealed in PDOS, there exists strong hybridization between Os's $5d$ and O's $2p$ orbitals around the Fermi energy level, owning to the spatial extended $5d$ orbitals. In fact, the previous neutron study also attributed the reduced moment to the hybridization between Os and O~\cite{Morrow:Prb}.

Furthermore, the same calculations have been done for La$_3$RuO$_7$ and the DOS/PDOS are shown in Fig.~\ref{Fig2}(c-d), which are qualitatively similar to La$_3$OsO$_7$. The local magnetic moment of Ru$^{5+}$ is $1.892$ $\mu_{\rm B}$ (SOC-enabled) or $1.900$ $\mu_{\rm B}$ (SOC-disabled), and such a negligible difference implies an weaker SOC effect comparing with La$_3$OsO$_7$. In particular, the magnitude of orbital moment is only $\sim0.018$ $\mu_{\rm B}$ per Ru, even lower than that of Os. The total moment is also lower than the ideal $3$ $\mu_{\rm B}$ but higher than the moment of Os, which is also reasonable considering the more localized distribution of $4d$ orbitals than $5d$. The reduced moment of Ru$^{5+}$ is also due to the covalency between Ru and O, as indicated in Fig.~\ref{Fig2}(c).

The calculated band gap of La$_3$RuO$_7$ is $0.70$ eV, which is higher than the experimental value ($\sim 0.28$ eV) extracted from transposrt~\cite{Khalifah:Prb1}. This inconsistent is probably due to their polycrystalline nature of samples and the presence of small amounts of the highly insulating La$_2$O$_3$, as admitted in Ref.~~\cite{Khalifah:Prb1}. More measurements, especially the optical absorption spectrum, are needed to clarify the intrinsic band gap of La$_3$RuO$_7$.

The aforementioned weak SOC effects to magnetism and band structures in La$_3$RuO$_7$ and La$_3$OsO$_7$ seem to contradict with the intuitive expection of strong SOC coefficients for $4d$/$5d$ electrons. This paradox can be understood as following. Since in La$_3M$O$_7$ the low-lying $t_{\rm 2g}$ orbitals are half-filled ($t_{\rm 2g}^3$), the Hund coupling between $t_{\rm 2g}$ electrons will prefer the high-spin state, in which the orbit moment is mostly quenched. Then the net effect of SOC is weak even if the SOC efficiency is large. Other $5d$ electronic systems with own more or less electrons than $t_{\rm 2g}^3$, e.g. Sr$_2$IrO$_4$, can active the SOC effects.

\subsection{Chemical doping and polaron forming}
Doping is a frequently used method to tune physical properties of materials. For Mott insulators, proper doping may result in superconductivity (e.g. for cuprates) or colossal magnetoresistivity (e.g. for manganites). One of the most anticipant doping effects on $5d$ metal oxides is the possible superconductivity, as predicted in Sr$_2$IrO$_4$~\cite{Wang:Prl, Watanabe:Prl}. However, till now, not only the superconductivity has not been found, but also there is an unsolved debate regarding the metallicity of doped Sr$_2$IrO$_4$. Some experiments reported the metallic transport behavior upon tiny doping and observed Fermi arcs using angle-resolved photoelectron spectroscopy (ARPES)~\cite{Kim:Np,Kim:Sci2,Torre:Prl,Korneta:Prb6}, while some others reported robust insulating (or semiconducting) behavior even upon heavy doping by element substitution and field-effect gating~\cite{Lu:Prb6,Calder:Prb6}.

Similarly, for La$_3$OsO$_7$, the experiment found that the Ca-doping up to $6.67\%$ could reduce the resistivity but the system remained insulating~\cite{Morrow:Prb}. Then it is interesting to investigate the doping effect. In our calculation, by using one Ca to replace one La in a unit cell, i.e. $8.33\%$ doping, the crystal structure is re-relaxed with various magnetism. Then the ground state turns to be AFM III, a little different from the original AFM IV (see Table 2 for more details). Even though, the in-chain AFM order remains robust.

\begin{table}
\centering
\caption{The energy difference (in unit of meV for $4$ f.u.'s) and local magnetic moments of $M$ in unit of $\mu_{\rm B}$ of doped La$_{11/4}$Ca$_{1/4}M$O$_7$. The SOC is disabled in calculations except for those items with "+SOC".}
\begin{tabular*}{0.73\textwidth}{@{\extracolsep{\fill}}llrc}
\hline
\hline
 $~$ & Magnetism & Energy & Moment \\
\hline
$M$=Os &      FM &     0 & 1.690,1.694,1.707,1.709 \\
                         &   AFM I &    79 & 1.559,1.622,-1.565,-1.219  \\
                         &  AFM II &   -63 & 1.647,1.645,-1.647,-1.699  \\
                   & AFM III &  -315 & 1.393,-1.393,-1.528,1.511  \\
       & AFM III (+SOC) & & 1.298,-1.298,-1.442,1.449  \\
                         &  AFM IV &  -285 & 1.257,1.617,-1.620,-1.289  \\
\hline
$M$=Ru &      FM &     0 & 1.753,1.765,1.797,1.803 \\
      &      FM (+SOC) &   & 1.749,1.753,1.791,1.792 \\
                         &   AFM I &   363 & 1.896,1.912,-1.887,1.265 \\
                         &  AFM II &     9 & 1.726,1.735,-1.789,-1.794 \\
                         & AFM III &    40 & -1.697,1.706,1.787,-1.749 \\
                         &  AFM IV &    42 & 1.568,1.891,-1.892,-1.531 \\
\hline
\hline
\end{tabular*}
\label{Table 2}
\end{table}

In contrast, when the $8.33$\% Ca-doping is applied to La$_3$RuO$_7$, our calculations predict that the ground state magnetism would probably transformed from AFM IV to FM, different from above Os-based counterpart. As summarized in Table 2, no matter the lowest energy FM state or the second lowest energy AFM II state, the in-chain FM order is unambiguous. This result is also reasonable considering the much weaker in-chain antiferromagnetism (i.e. $J_1$) of La$_3$RuO$_7$. Thus the antiferromagnetism of La$_3$RuO$_7$ should be more fragile against chemical doping. Further experiments are needed to verify our prediction.

\begin{figure}
\centering
\includegraphics[width=0.68\textwidth]{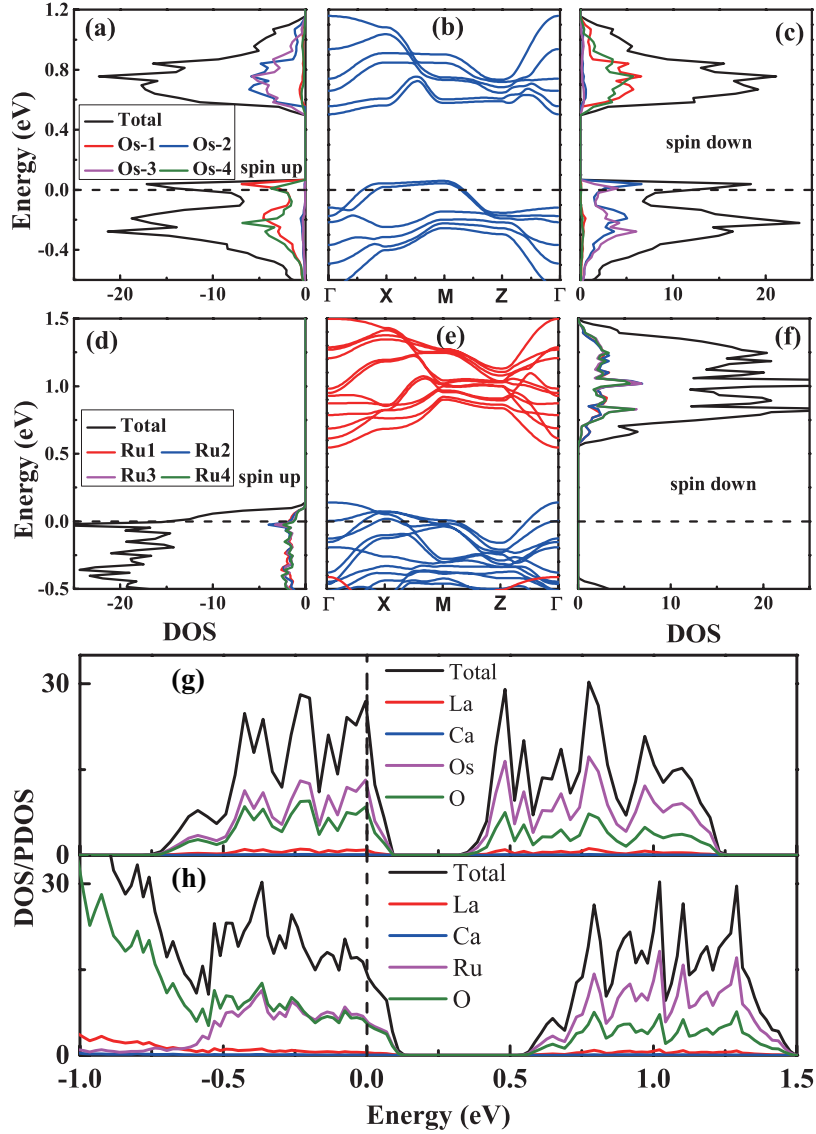}
\caption{DOS/PDOS and corresponding band structures around the Fermi level of La$_{11/4}$Ca$_{1/4}M$O$_7$ ($M$=Os or Ru). (a-c) La$_{11/4}$Ca$_{1/4}$OsO$_7$; (d-f) La$_{11/4}$Ca$_{1/4}$RuO$_7$. In (e), spin up/down bands are distinguished by blue/red colors. (g-h) The SOC-enabled DOS/PDOS which are only slightly changed compared with the non-SOC ones.}
\label{Fig3}
\end{figure}

As shown in Fig.~\ref{Fig3}, the DOS's of doped La$_3$OsO$_7$ and La$_3$RuO$_7$ own finite values at Fermi levels, implying metallic behavior, which seems to be opposite to experimental observation of doped La$_3$OsO$_7$. However, a careful analysis finds that this finite DOS at Fermi level should be due to a technical issue of calculation. The substitution of one La by one Ca will bring one hole into the system. However, the AFM state implies at least doubly degenerate bands (spin up and spin down). Thus one hole to doubly degenerate bands always leads to half-filling, as observed in our DOS. Thus, the finite DOS at Fermi level does not guarantee metallicity of Ca-doped La$_3$OsO$_7$, while the metallicity of Ca-doped La$_3$RuO$_7$ needs experimental verification.

The SOC-enabled calculations have also been performed for the doped La$_3M$O$_7$. However, due to the partial hole concentration ($\sim1/4$ per $M$), the SOC effect is not prominent. For example, the near-Fermi-level DOS (Fig.~\ref{Fig3}(g-h)) are similar to the corresponding non-SOC ones. The local moments for the ground states are also listed in Table~\ref{Table 2}, which are only slightly lower than the original one without SOC, especially for the Ru case.

The PDOS's of Ca-doped La$_3$OsO$_7$ show that the Os ions can be classified into two types: a) two near-Ca Os's (one spin up and one spin down); b) other two Os's. Their PDOS's are slightly different, and the type-b Os's are less affected by the Ca-doping, namely the doping effect has a tendency to be localized. In contrast, the PDOS's of Ca-doped La$_3$RuO$_7$ show that almost all four Ru ions are equally effected by the Ca-doping.

\begin{figure}
\centering
\includegraphics[width=0.48\textwidth]{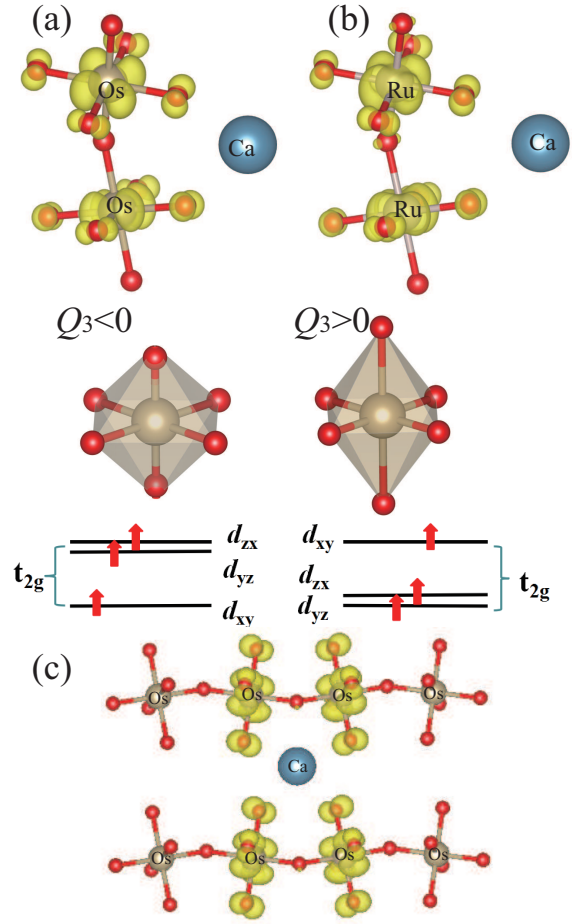}
\caption{Plots of charge density distribution of hole in Ca-doped La$_3$OsO$_7$ (a) and La$_3$RuO$_7$ (b). Lower: schematic of corresponding Jahn-Teller $Q_3$ mode. In La$_3$OsO$_7$, $Q_3$ is negative, leading to a low-lying $d_{xy}$ orbitals for electron. In La$_3$RuO$_7$, $Q_3$ is positive, leading to a higher-energy $d_{xy}$ orbitals for electron. However, the orbital shapes of hole after Ca-doping are very similar between these two systems,  suggesting the Coulombic interaction between hole and dopant to be the common driven force. (c) The spatial hole distribution in Ca-doped La$_3$OsO$_7$ supercells doubled along the $c$ axes. The corresponding doping concentration is $4.17\%$. Similar results are obtained for Ca-doped La$_3$RuO$_7$ supercells (not shown).}
\label{Fig4}
\end{figure}

To clarify the effect of doping, the charge density distribution of hole in Ca-doped La$_3$OsO$_7$ is visualized in Fig.~\ref{Fig4}(a). Here we only extract the wave function of the above-Fermi-level partial bands, which can represent the hole (half spin-up hole plus half spin-down hole). Clearly, the orbitals of hole are the $d_{xy}$ type on Os site and the $p_x$ type on O site (the chain direction is chosen as the $z$ axis), implying an spatially extended wave function. According to the Slater-Koster equation~\cite{Slater:Pr}, the lying-down $d_{xy}$ orbital has a very weak hopping amplitude along the $z$-axis, if not ideally zero. Thus, the hole will be restricted near the Ca dopant by the Coulombic interaction, leading to the semiconducting behavior of Ca-doped La$_3$OsO$_7$.

According to the plenty experience of $3d$ electron systems, the lattice distortions, e.g. Jahn-Teller modes, will be always activated by partially occupied $t_{2g}$ orbitals (or $e_g$ orbitals) to split the energy degeneration between/among orbitals. By carefully analyzing the bond lengths of oxygen octahedra, it is easy to verify the effect of hole modulated lattice distortions. First, the breathing mode $Q_1$ can be defined as $(l_x+l_y+l_z)/\sqrt{3}$ to characterize the size of oxygen octahedral cage, where $l$ denote the O-$M$-O bond length along a particular axis~\cite{Dagotto:Prp,Dong:Prb11}. After the doping, the changes of $Q_1$ are $-5.307$ pm for the near-Ca Os's and $-3.636$ pm for other two Os's. These shrunk octahedral cages are due to the Coulombic attraction between positive-charged hole on Os and negative charged oxygen ions. Second, the Jahn-Teller modes $Q_2$ and $Q_3$ can be defined as $(l_x-l_y)/\sqrt{2}$ and $(-l_x-l_y+2l_z)/\sqrt{6}$ respectively, which can split the degeneration among triplet $t_{2g}$ orbitals or between doublet $e_g$ orbitals. For La$_3$OsO$_7$, the original Jahn-Teller modes $Q_2$=$0.100$ pm and $Q_3$=$-3.938$ pm. This prominent $Q_3$ mode prefers the $d_{xy}$ orbital for electrons. Therefore, the $d_{xy}$ hole after doping is \emph{not} driven by this preseted lattice distortion, but can only be due to the Coulombic interaction from Ca$^{2+}$ since the spatial distribution of $d_{xy}$ hole is more closer to dopant (see Fig.~\ref{Fig4}). Then this Coulombic-driven $d_{xy}$ hole will suppress the $Q_3$ mode: $Q_3$=$-2.414$ pm for near-Ca Os's and $-1.007$ pm for other two Os's. Meanwhile, the $Q_2$ mode are enhanced: $Q_2$=$0.942$ pm for near-Ca Os's and $1.152$ pm for the other two Os's.

For doped La$_3$RuO$_7$, the changes of $Q_1$ are $-4.074$ pm for the near-Ca Ru's and $-3.470$ pm for the other two, similar to the case of doped La$_3$OsO$_7$. For original La$_3$RuO$_7$, the Jahn-Teller modes are: $Q_2$=$0.110$ pm, $Q_3$=$5.351$ pm. In contrast with La$_3$OsO$_7$, this lattice distortion dislike the $d_{xy}$ orbital (for electron) in energy. Then the Coulombic-driven $d_{xy}$ hole will further enhance this positive $Q_3$ mode: $Q_3$=$7.574$ pm for the near-Ca Ru's and $7.913$ pm for the other two Ru's. Meanwhile, the $Q_2$ mode are enhanced as in La$_3$OsO$_7$: $Q_2$=$1.227$ pm for the near-Ca Ru's and $1.129$ pm for the other two.

The localization of hole can be further confirmed by calculations of supercells. As shown in Fig.~\ref{Fig4}(c-e), the hole occupancies on four near-Ca Os ions are much prominent than other Os ions which stay only one u.c. space from the dopant. Similar results exist for doped La$_3$RuO$_7$. This localized hole with distorted lattices form the polaron.

According previous studies, there are magnetic polarons in manganites, which are ferromagnetic clusters of several Mn sites embedded in AFM background~\cite{Meskine:Prl}. This scenario is quite possible for La$_3$RuO$_7$ considering the ferromagetically-aligned Ru ions as revealed in Table~\ref{Table 2}. However, for La$_3$OsO$_7$, both the result of the minimal cell (shown in Table~\ref{Table 2}) and the calculation of doubled-supercell (along the $c$-axis) dislike magnetic polaron, at least the small (up to three-site) magnetic polaron. It is also reasonable considering the differences among $3d$/$4d$/$5d$ electrons. The $3d$ electrons have strong Hubbard interaction which prefers localized magnetic moments. Thus the energy gain from forming magnetic polaron is large~\cite{Meskine:Prl}. In contrast, the $5d$ electrons are more spatially extended and own weaker Hubbard interaction, which are disadvantage to form magnetic polaron. In fact, in Ref.~~\cite{Meskine:Prl}, the second nearest-neighbor hopping (which equals to the spatially extended effect) can suppress the formation of magnetic polaron, which provides a hint to understand the difference between $3d$ polaron and $5d$ polaron. Of course, the current calculations can not fully exclude the possibility of magnetic polarons with larger sizes or higher dimensional, which are beyond our computational capability considering the fact that a minimal cell of La$_3$OsO$_7$ already contains $44$ ions.

In short, considering above results of hole restricted by Coulombic interaction and the lattice distortions followed the particular $d_{xy}$-orbital hole, it is reasonable to argue that the carriers generate by Ca doping would be mostly localized near dopant to form polaron and contribute to the semiconducting behavior. This scenario may explain the puzzle why the expected metallicity is absent in some doped $5d$-Mott-insulators.

\section{Conclusion}
In summary, two $4d$/$5d$ metal oxides La$_3M$O$_7$ ($M$=Os and Ru) with unique quasi-one-dimensional $M$ chains have been studied systematically using the density function theory calculation. Their magnetic ground states are revealed to be identical, in agreement with the recent neutron study of La$_3$OsO$_7$ but different from early calculations on La$_3$RuO$_7$. Due to the half-filled $t_{2g}$ configuration, the spin-orbit coupling is not crucial in these two systems. Moreover, the doping of Ca has been predicted to affect the magnetism somewhat, leading to different effects for $M$=Os and Ru. In particular, the $d_{xy}$-orbital-sharp hole is formed after Ca dopant. This orbital is driven by the Coulombic interaction, and further tunes the lattice distortions. The experimental observed semiconducting behavior after doping is explained as the forming of polarons.

\section*{Acknowledgment}
This work was supported by National Natural Science Foundation of China (Grant No. 11674055) and Fundamental Research Funds for the Central Universities.

\section*{References}
\bibliography{ref}
\end{document}